\documentstyle[12pt]{article} 

\def\a{\alpha}
\def\b{\beta}
\def\c{\chi}
\def\d{\delta}
\def\e{\epsilon}
\def\ve{\varepsilon}
\def\f{\phi}
\def\vf{\varphi}
\def\g{\gamma}

\def\k{\kappa}

\def\m{\mu}
\def\n{\nu}
\def\o{\omega}
\def\p{\pi}

\def\r{\rho}
\def\s{\sigma}
\def\t{\tau}

\def\x{\xi}
\def\z{\zeta}

\def\Ld{\Lambda}

\def\S{\Sigma}

\def\cf{{\cal F}}

\def\inbar{\vrule height1.5ex width.4pt depth0pt}
\def\rlx{\relax\leavevmode}
\def\I{\leavevmode\hbox{\small1\kern-3.8pt\normalsize1}}
\def\openone{\leavevmode\hbox{\small1\kern-3.3pt\normalsize1}}
\def\Ione{\rlx{\rm 1\kern-2.7pt l}}
\font\cmss=cmss10
\font\cmsss=cmss10 at 7pt
\def\ZZ{\rlx\leavevmode
             \ifmmode\mathchoice
                    {\hbox{\cmss Z\kern-.4em Z}}
                    {\hbox{\cmss Z\kern-.4em Z}}
                    {\lower.9pt\hbox{\cmsss Z\kern-.36em Z}}
                    {\lower1.2pt\hbox{\cmsss Z\kern-.36em Z}}
               \else{\cmss Z\kern-.4em Z}\fi}
\def\Ik{\rlx{\rm I\kern-.18em k}}  % Yes, I know. This ain't capital.
\def\IC{\rlx\leavevmode
             \ifmmode\mathchoice
                    {\hbox{\kern.33em\inbar\kern-.3em{\rm C}}}
                    {\hbox{\kern.33em\inbar\kern-.3em{\rm C}}}
                    {\hbox{\kern.28em\sinbar\kern-.25em{\rm C}}}
                    {\hbox{\kern.25em\ssinbar\kern-.22em{\rm C}}}
             \else{\hbox{\kern.3em\inbar\kern-.3em{\rm C}}}\fi}
\def\IP{\rlx{\rm I\kern-.18em P}}
\def\IR{\rlx{\rm I\kern-.18em R}}
\def\IN{\rlx{\rm I\kern-.20em N}}

\def\llsymbol#1{\@llsymbol{\@nameuse{c@#1}}}
\def\@llsymbol#1{\ifcase#1\or {}\or {'}\or {''}\or {'''}\or
   {''''}\or {'''''}\or  \else\@ctrerr\fi\relax}

\newcounter{contador}
\newcommand{\letra}{
   \stepcounter{equation}
   \setcounter{contador}{\value{equation}}
   \setcounter{equation}{0}
   \renewcommand{\theequation}{\thecontador.\alph{equation}}}
\newcommand{\antiletra}{
   \renewcommand{\theequation}{\arabic{equation}}
   \setcounter{equation}{\value{contador}}}

\newcommand{\ol}\overline
\newcommand{\ti}\tilde
\newcommand{\wt}\widetilde
\newcommand{\wh}\widehat
\newcommand{\bv}\breve
\newcommand{\dg}\dagger
\newcommand{\dr}{\stackrel{\mbox{\scriptsize DR}}\longrightarrow}

\newcommand{\aand}{\;\;\;\mbox{and}\;\;\;}

\newcommand{\be}{\begin{equation}}
\newcommand{\ee}{\end{equation}}
\newcommand{\bl}{\begin{eqnarray}&}
\newcommand{\el}{&\end{eqnarray}}
\newcommand{\bq}{\begin{eqnarray}}
\newcommand{\eq}{\end{eqnarray}}

\newcommand{\sss}{\scriptscriptstyle}
\newcommand{\dst}{\displaystyle}

\newcommand{\sm}{{\s}_{\m}}
\newcommand{\sn}{{\s}_{\n}}

\newcommand{\ov}{\overline}
\newcommand{\pa}{\partial}

\newcommand{\ab}{\bar A}
\newcommand{\bb}{\bar B}
\newcommand{\ats}{\bar s}
\newcommand{\atd}{\bar\d}
\newcommand{\atc}{\bar c}
\newcommand{\BRS}{$\ov{\mbox{BRS}}$}
\newcommand{\az}{\bar\z}
\begin{document}
\begin{titlepage}
\topskip0.5cm
\hfill\hbox{TUW--98-14}

\hfill\hbox{hep-th/9806137}\\[2cm]
\begin{center}{\Large\bf
Twisted $N=4$ SUSY Algebra in Topological Models of Schwarz Type\\
[2cm]}{\large
O. M. Del Cima\footnote{Work supported by the {\it Fonds zur 
F\"orderung der Wissenschaftlichen Forschung (FWF)} under the contract 
number P11654-PHY, e-mail: delcima@tph73.tuwien.ac.at.},
K. Landsteiner\footnote{Work supported by the {\it Fonds zur 
F\"orderung der Wissenschaftlichen Forschung (FWF)} under the contract 
number P10268-PHY, e-mail: landstei@tph45.tuwien.ac.at.},
M. Schweda\footnote{e-mail: schweda@tph.tuwien.ac.at.}\\[1.2cm]}
Institut f\"ur theoretische Physik\\
TU-Wien\\
Wiedner Hauptstra{\ss}e 8-10\\ 
A-1040 Wien, Austria
\end{center}
\vskip2.cm
\begin{abstract}
We reinvestigate the twisted $N=4$ supersymmetry present in
Schwarz type topological field models. We show that Chern-Simons
theory in three dimensions can be untwisted to a kind of sigma-model 
with reversed statistics only in the free case.
By dimensional reduction we define then the two-dimensional BF-model.
We establish an analog result concerning the untwisting. As a consequence
of the definition through dimensional reduction we find new fermionic
scalar symmetries that have been overlooked so far in the literature.
\end{abstract}
\vfill
\hbox{June 1998}\hfill\\
\end{titlepage}

%%%%%%%%%%%%%%%%%%%%%%%%%%
\section{Introduction}
%%%%%%%%%%%%%%%%%%%%%%%%%%
%\vspace{3cm}

Topological field models have been a subject of high interest in the 
last years.
One of the highlights is the presence of a particular kind of supersymmetry.
The supercharges in the conventional case transform as spinors under the
rotation group. In topological models however, they transform as scalar and
vectors, giving rise to an algebra of the form
\be
\{s,\delta_\m\} = \partial_\m\,.
\label{topsusy}
\ee  
This is usually called topological supersymmetry. 

There are two different kinds of topological field theories. The first one is
called Witten-type, and the prime example is topological Yang-Mills theory in
four dimensions \cite{witten}. Here one starts with $N=2$ super Yang-Mills 
theory. The $N=2$ 
superalgebra has an $SU(2)_{left}\otimes SU(2)_{right}\otimes SU(2)_I$ 
automorphism algebra. The rotation group $SO(4)$ is given by the product of
the first two 
$SU(2)$ factors and $SU(2)_I$ is an internal symmetry. The supercharges 
transform as $(\frac 1 2 , 0, \frac 1 2)$
and $(0,\frac 1 2, \frac 1 2 )$. In the topologically twisted theory one 
takes as a new rotation group the diagonal subgroup of $SU(2)_{left}\otimes 
SU(2)_I$ times $SU(2)_{left}$.
The supercharges transform as $(0,0)$, $(3,0)$ and $(2,2)$ under this 
new rotation group. The singlet is a fermionic, nilpotent symmetry and 
constitutes a part of the BRS-operator of the theory. Together with the 
vector  \cite{bmops} it gives rise to an algebra of the form (\ref{topsusy}).

The other kind of topological field theory is called Schwarz-type. Here one 
starts with a Lagrangian that is an $n$-form, such that it can be integrated 
over a $n$-dimensional manifold without the need of a metric. 
Examples are Chern-Simons theories in $(2m+1)$ dimensions and BF-models
in arbitrary dimensions. 
The Lagrangian 
typically has a gauge symmetry. So upon quantization one is forced to 
introduce a gauge fixing {\it \`a la} BRS. The action is written as
an invariant part plus the gauge fixing part. The gauge fixing part depends
on the metric and can be written as a BRS-variation. 
It turns out then, that the gauge 
fixing 
procedure gives rise to a fermionic vector symmetry that together
with the BRS-operator obeys an algebra of the form (\ref{topsusy}). This
vector supersymmetry has also played an important role in the algebraic 
study of field theories \cite{pigsor}. In particular it
has given rise {\it e.g.} to a finiteness prove of 
Chern-Simons theory \cite{finiteness}.

In Schwarz-type topological models the algebra (\ref{topsusy}) does not come 
through a twisted supersymmetric structure but rather expresses the 
triviality of translation as a consequence of the topological nature of the
theory.
Nevertheless it can sometimes be completed to an extended (twisted) 
supersymmetry. We will demonstrate this on the example of Chern-Simons 
theory. Using the fact that the gauge fixing term can not only be written as 
a BRS-variation but also as an anti-BRS-variation we derive a twisted $N=4$ 
algebra in gauge fixed Chern-Simons theory. With somewhat different methods 
this result has been established also in \cite{silvio1}. Then we investigate 
the possibility of defining Chern-Simons theory as a topological field theory 
of Witten-type\footnote{Often the action of Witten-type models can be 
written as a BRS-variation. This is not always so as the example of the
twisted sigma-model in three dimension as defined in \cite{roswit} shows.}. 
This means that we look for a kind of sigma-model action 
whose topologically twisted version is gauge fixed Chern-Simons theory. We 
find that this is possible only in the case with vanishing interaction
terms. The two-dimensional BF-model is then defined through dimensional
reduction from the Chern-Simons action. In this way we are also able to 
establish a twisted $N=4$ supersymmetry for this model. As part of this we 
find two new fermionic, nilpotent (pseudo-)scalar symmetries, that have been 
overlooked so far in the literature. We also investigate the question of 
untwisting two-dimensional BF-model finding again that it is possible only 
in the non-interacting case.

%%%%%%%%%%%%%%%%%%%%%%%%%%%%%%%%%%%%%
\section{The $N=4$ supersymmetry algebra in $D=3$}
%%%%%%%%%%%%%%%%%%%%%%%%%%%%%%%%%%%%%
The $N=4$ supersymmetry algebra in Euclidean 
$D=3$\footnote{For conventions and notations adopted throughout the
work see the Appendix.}, assuming {\it a priori} the absence of central 
charges, is given by :
\be
\{Q_{\a}^{~A\ab},Q_{\b}^{~B\bb}\}=2i\e^{AB}\e^{\ab\bb}\s_{\m\a\b}\pa_\m ~~.
\label{tn4}
\ee
%where the isospin indices are labeled by the overlined capital letters 
%($\ab,\bb$$=$$+,-$), all the spinor indices are labeled from now on by the 
%first letters of the Greek alphabet ($\a,\b,\g,\d$$=$$1,2$) -- even for the 
%capital letters $A$ and $B$, since after the twisting, they have the same 
%status as the spinor indices $\a$ and $\b$ -- and the Euclidean space 
%indices by the Greek letters ($\m,\n,\r,\s$$=$$1,2,3$). Therefore,
%the algebra (\ref{tn4}) can be written as
%\be
%\{Q_{\,\a}^{\ab~\g},Q_{\,\b}^{\bb~\d}\}=2i\e^{\g\d}\e^{\ab\bb}
%\sm_{\a\b}\pa_\m ~~.
%\label{tn41}
%\ee
On this algebra acts an $SU(2)_{Spin}\otimes SU(2)_I \otimes SU(2)_{II}$.
The supercharges transform as $(\frac 1 2 ,\frac 1 2,\frac 1 2)$. 
Spinor indices are denoted by Greek letters ($\a,\b,\g,\d=1,2$), and 
the isospin indices
for $SU(2)_I$ by capital Latin letters, the ones for $SU(2)_{II}$ by
capital, barred Latin letters. The algebra can be obtained by dimensional
reduction of the $N=2$ algebra in four dimensions. Let us describe now how
the topological twist acts. We chose as the new rotation group the 
diagonal subgroup of $SU(2)_{Spin}\otimes SU(2)_I$. For practical purposes
this is done by identifying the capital Latin indices with the Greek indices
$A \rightarrow \a, \cdots$. So the indices for the new rotation group
are also denoted by Greek letters. The Euclidean space indices 
are denoted by the Greek letters ($\m,\n,\r,\s=1,2,3$).
When we decompose the generators of (\ref{tn4}) into irreducible 
representations of the new rotation group with can define the 
operators $s^{\ab}$ and $\d^{\ab}_{\m}$ :
\letra
\bq
&&s^{\ab}=\left(\begin{array}{c}
s\\
\ats
\end{array}\right)\equiv {1\over{2}}~\d_\a^{~\b}Q_{\,\b}^{\ab~\a}
~~~{\mbox{(2 d.f. from Tr$(Q^{\ab})$)}} ~~;\label{brs} \\
&&\d^{\ab}_{\m}=\left(\begin{array}{c}
\atd_\m\\
\d_\m
\end{array}\right)\equiv -{i\over{2}}~\s_{\m\a}^{~~\b}Q_{\,\b}^{\ab~\a} 
~~~{\mbox{(6 d.f. from Tr$(\s_\m Q^{\ab})$)}}~~.\label{topvec}
\eq
\antiletra
The definitions given above (eqs.(\ref{brs}--\ref{topvec})) together with the 
algebra (\ref{tn4}) (considering the conventions presented in the Appendix), 
yield the anticommutation relations displayed below :
\be
\{s^{\ab},s^{\bb}\}=0 ~~,~~\{s^{\ab},\d^{\bb}_{\m}\}=-\e^{\ab\bb}\pa_\m \aand 
\{\d^{\ab}_{\m},\d^{\bb}_{\n}\}=\e^{\ab\bb} \ve_{\m\n\r}\pa_\r ~~,
\label{alg1}
\ee 
which give the following algebra :
\bq
&&\{s,s\}=\{\ats,s\}=\{s,\ats\}=\{\ats,\ats\}=0 ~~;\nonumber \\
&&\{s,\d_\m\}=\pa_\m ~~,~~\{s,\atd_\m\}=0 ~~;\nonumber \\
&&\{\ats,\d_\m\}=0 ~~,~~\{\ats,\atd_\m\}=-\pa_\m ~~;\nonumber \\
&&\{\d_\m,\d_\n\}=0 ~~,~~\{\d_\m,\atd_\n\}=\ve_{\m\n\r}\pa_\r ~~;\nonumber \\
&&\{\atd_\m,\d_\n\}=-\ve_{\m\n\r}\pa_\r ~~,~~\{\atd_\m,\atd_\n\}=0 ~~.
\label{alg2}
\eq

As it will be shown afterwards, the algebra (\ref{alg2}) fulfilled by the 
operators $s$, $\ats$, $\d_\m$ and $\atd_\m$, is exactly the same 
satisfied by the 
BRS and the anti-BRS (\BRS) operators, together with their respectives 
topological 
vector supersymmetries, in the case of the Chern-Simons model in a 
three-dimensional Euclidean space.

Now, it is possible to write the supercharges $Q_{\,\a}^{\ab~\b}$ in terms of 
$s$, $\ats$, $\d_\m$ and $\atd_\m$. By solving the equations (\ref{brs}) and 
(\ref{topvec}), it is easily shown that the generators 
$Q_{\,\a}^{+~\b}$ and $Q_{\,\a}^{-~\b}$ read
\letra
\bq
&&Q_{\,\a}^{+~\b}=\d_\a^{~\b}s+i\s_{\m\a}^{~~\b}\atd_\m
~~\Longrightarrow~~ Q^+=\I s+i\sm\atd_\m ~~,\label{Q+}\\
&&Q_{\,\a}^{-~\b}=\d_\a^{~\b}\ats+i\s_{\m\a}^{~~\b}\d_\m
~~\Longrightarrow~~ Q^-=\I\ats+i\sm\d_\m ~~.
\label{Q-}
\eq
\antiletra
In a further Section it will be seen that, in the case of Chern-Simons, 
the supersymmetry charges $Q^+$ and $Q^-$ have 
%eingenvalues 
%$+1$ and $-1$, respectively. These quantum numbers are associated to 
the ghost numbers $+1$ of the BRS operator ($s$) and $-1$ of the 
\BRS~operator ($\ats$). Therefore, it means that, in the case of Chern-Simons, 
these supersymmetry generators relates fields (in a non-linear way) 
with ghosts numbers differed by
a unity. It should be stressed that the Chern-Simons model has actually a
 $N=4$ supersymmetry, but in this case the ``superpartners'' are 
not related anymore by a half spin, in fact, they are related by a unity of
ghost number.
 
%%%%%%%%%%%%%%%%%%%%%%%%%%%%%%%%%%%%%
\section{The operators $Q^+$ and $Q^-$ in topological models}
%%%%%%%%%%%%%%%%%%%%%%%%%%%%%%%%%%%%%
In this Section, we consider the inverse problem by supposing  
a topological model (Chern-Simons, as a particular case) which the BRS, the
\BRS~and their respectives topological symmetries satisfy the algebra
given by eq.(\ref{alg2}).

\subsection{The inverse problem}
Assuming the algebra (\ref{alg2}) and defining the $Q^+$ and $Q^-$ operators
as in eqs.(\ref{Q+}--\ref{Q-}), it can be easily checked that their 
anti-commutation relations read
\letra
\bq
&&\{Q^+,Q^+\}=0 \aand \{Q^-,Q^-\}=0~~;\\ 
&&\{Q^+,Q^-\}=\left[i(\I\wedge\sm)\pa_\m + 
{1\over 2}(\sm\wedge\sn)\ve_{\m\n\r}\pa_\r\right]=
-2i\e^{\g\d}\s_{\m\a\b}\pa_\m~~, \\
&&\{Q^-,Q^+\}=-\left[i(\I\wedge\sm)\pa_\m + 
{1\over 2}(\sm\wedge\sn)\ve_{\m\n\r}\pa_\r\right]=
2i\e^{\g\d}\s_{\m\a\b}\pa_\m~~, \label{n4inv}
\eq
\antiletra
where the $N=4$ supersymmetry charges $Q^+$ and $Q^-$ are 
nilpotent operators with ghost numbers $+1$ and $-1$, respectively. Therefore,
a two new nilpotent operators appear, in a natural way, as a by-product of 
the topological nature of a model and its $N=4$ supersymmetry.

\subsection{The Chern-Simons model}
Here, our goal, is to show that the Chern-Simons model has a $N=4$ 
supersymmetry as a consequence of being a topological model, as well 
as, to find its untwisted version with the help of the nilpotent operators 
(\ref{Q+}--\ref{Q-}).

The Chern-Simons action is given by 
\be
\S_{{\rm CS}}={k\over 4\p}{\mbox{Tr}}\int d^3x~\ve
_{\m\n\r}\left( A_\m\pa_\n A_\r+{2\over 3}A_\m A_\n A_\r\right) ~~,
\label{CS}
\ee
where the parameter $k$ is quantized~\cite{witten} and $A_\m$ is a Lie
algebra valued gauge field, as well as all the fields we shall introduce
later on. Denoting such a generic field by $\f$, we define
\be
\f(x)\equiv\f^a(x)\t_a ~~, 
\ee
where the matrices $\t$ are the generators of the group and obey
\be
[\t_a,\t_b]=f_{abc}\t_c \aand {\mbox{Tr}}(\t_a\t_b)={1\over 2}\d_{ab} ~~.
\ee
The gauge transformations read
\be
\d A_\m=-(\pa_\m\o + [A_\m,\o])\equiv -D_\m\o ~~,
\ee
with $\o\equiv\o^a\t_a$. The gauge group is chosen to be simple and compact. 
These transformations change the integrand of the action (\ref{CS}) into a 
total derivative, leaving thus the action invariant if there are no boundary 
contributions and if the topology is trivial~\cite{witten}. 

The gauge fixing is of the Landau-type. It is implemented by adding to the 
gauge invariant action (\ref{CS}) the term
\be
\S_{{\rm gf}}={\mbox{Tr}}\int d^3x~
\left( b~\pa_\m A_\m + \atc~\pa_\m D_\m c \right) ~~,\label{gf}
\ee
where the Lie algebra valued fields $c$, $\atc$ and $B$, are the ghost, the 
antighost and a Lagrange multiplier, respectively.

The gauge fixing action (\ref{gf}) can also be written either as a 
BRS-variation or as a \BRS-variation :
\bq
\S_{{\rm gf}}=
&\left\{\begin{array}{l}
s~{\mbox{Tr}}{\dst\int} d^3x~\left(\atc~\pa_\m 
A_\m\right) \\
\\
-\ats~{\mbox{Tr}}{\dst\int} d^3x~\left(c~\pa_\m A_\m\right) 
~~~~. \label{sgf}
\end{array}\right.
\eq
Therefore, the action 
$\S_{{\rm inv}}^{{\rm (3)}}=\S_{{\rm CS}}+\S_{{\rm gf}}$ 
is invariant under the following BRS and \BRS~transformations :
\be
\begin{array}{ll}
sA_\m=-D_\m c ~~,& \ats A_\m=-D_\m \atc ~~,\nonumber\\
sc=c^2 ~~,& \ats\atc=\atc^2 ~~,\nonumber\\
s\atc=b ~~,& \ats c=\{\atc,c\}-b ~~,\nonumber\\
sb=0 ~~,& \ats b=[\atc,b] ~~.\label{transf} 
\end{array}
\ee

Since the action $\S_{{\rm inv}}^{{\rm (3)}}$ depends on the metric only 
through the gauge fixing part (\ref{sgf}), which is a BRS or a \BRS-variation,
the energy-momentum tensor will be a BRS or a \BRS-variation as well : 
\bq
\left. T_{\m\n}={2\over{\sqrt g}}
{{\d\S_{\rm gf}}\over{\d g_{\m\n}}}\right|_{g_{\m\n}=\d_{\m\n}}=
&\left\{\begin{array}{l}
s \Ld_{\m\n} \\
\\
\ats {\bar\Ld}_{\m\n}
~~~. \label{tmn}
\end{array}\right.
\eq
Beyond the original topological vector supersymmetry ($\d_\m$) associated
to the BRS symmetry~\cite{vecsusy,silvio1}, from eq.(\ref{tmn}), we can 
infer about the existence of another topological vector supersymmetry 
associated to the \BRS~symmetry ($\atd_\m$), since, due to the gauge fixing 
(\ref{sgf}), the energy-momentum tensor can also be written as a 
\BRS-variation. 
By using the same approach to the usual 
case\footnote{For details concerning the topological vector supersymmetry 
see~\cite{pigsor,olivier,algvecsusy} and references therein.}\cite{pigsor}, it 
can be found that the action $\S_{{\rm inv}}^{{\rm (3)}}$ is
invariant under the following topological vector supersymmetries :
\be
\begin{array}{ll}
\d_\m A_\n={2\p\over{k}}\ve_{\m\n\r}\pa_\r\atc ~~,& 
\atd_\m A_\n={2\p\over{k}}\ve_{\m\n\r}\pa_\r c ~~,\nonumber\\
\d_\m c=-A_\m ~~,& \atd_\m\atc=A_\m  ~~,\nonumber\\
\d_\m b=\pa_\m\atc ~~,& \atd_\m b=D_\m c ~~,\nonumber\\
\d_\m\atc=0 ~~,& \atd_\m c=0 ~~.\label{vecsusyt} 
\end{array}
\ee

It should be stressed that the BRS ($s$) and the 
\BRS~($\ats$), eq.(\ref{transf}), together with their respectives topological 
vector 
supersymmetries ($\d_\m$) and ($\atd_\m$), eq.(\ref{vecsusyt}), 
form the algebra given by 
eq.(\ref{alg2}), which closes up to contact-terms.

\subsection{Untwisting Chern-Simons}
Given the fact that the Chern-Simons model with Landau-type
gauge fixing posses a twisted $N=4$ supersymmetry it is tempting
to speculate if it can be defined as a topological model of Witten-type.
This would imply that there exists a kind of sigma-model with $N=4$
supersymmetry, whose twisted version would be the gauge-fixed Chern-Simons
action. Actually it would be a sigma-model of reversed statistics, since
the scalar fields should be related to the ghosts of the Chern-Simons theory.
On the other hand the spinor fields would be bosons since they should,
after twisting, give the gauge field $A_\mu$. Such a spinor field can
be defined using the bosonic fields of the gauge-fixed Chern-Simons model
as follows \footnote{We think of the
fields corresponding to differentialforms as being complexified. 
This is necessary since in three dimensions with Euclidean-signature we
do not have Majorana spinors.
As explained in \cite{marcos} going from the untwisted to the twisted 
form of the Lagrangian also includes a restriction to real fields. Similarly,
after defining the (untwisted) spinor fields we could go to Minkowski-signature
and declare the spinors to be Majorana. Here we take the twist as a rather
formal manipulation and ignore these subtelties in the following.}:
\bq
\c_{\a}^{~A}={1\over 2}\left(\d_\a^{~A} \sqrt{\frac{2\p}{k}} b-i\s_{\m\a}^{~~A}
\sqrt{\frac{k}{2\p}}A_\m\right)\,,\nonumber\\
b=\sqrt{\frac{k}{2\p}}{\mbox{Tr}}(\c) \aand A_\m=i\sqrt{\frac{2\p}{k}}
{\mbox{Tr}}(\s_{\m}\c) ~~.
\label{x}
\eq
According to our index conventions it is made explicit that we take
$\c$ to transform as $(\frac 1 2 ,\frac 1 2 )$ under the $SU(2)_{Spin}\otimes
SU(2)_I$ whereas $b$ and $A_\mu$ transform as a singlet and a triplet
under the diagonal subgroup. Of course this is precisely how the
topological twist works in three dimensions. Let us study now a
kinetic term making the $SU(2)_{Spin}\otimes SU(2)_I\otimes SU(2)_{II}$
structure of the $N=4$ algebra explicit. Demanding invariance under
this product group fixes the kinetic terms to be
\be
S_{\rm kin}={\mbox{Tr}}\int d^3x~\left(i\e_{AB}~\c_\a^A\s_\m^{\a\b}
\pa_\m\c_\b^B + {\frac 12} \e_{\ab\bb}c^{\ab} \pa^2 c^{\bb}\right)~~,
\label{kinetic}
\ee
where we defined $c^{\ab}=(\atc, c)$.
Using (\ref{x}) one sees that this indeed reproduces the terms containing
derivatives of the Chern-Simons action. In particular we also obtain the
gauge fixing term $b\pa_\m A_\m$ ! Non-Abelian Chern-Simons theory
also contains interaction terms. The question is now if one can also 
obtain these Chern-Simons interactions from the twisted version of some
interactions between the spinors $\c_\a^A$. From the form of the interaction
\be
S_{\rm int}={\mbox{Tr}}\int d^3x~\left[\frac{k}{6\p} \ve_{\m\n\r} 
A_\m A_\n A_\r + \atc~\pa_\m\left([A_\m,c]\right)\right]
\label{int}
\ee
it is clear that one will need the product
of three $\c_\a^A$'s to produce the term with three gauge fields. 
Let us concentrate now on the $SU(2)_{Spin}$-group,
under which $\c$ transforms as spin $\frac 1 2$ representation. The product
of three of these fields transforms as $\frac 1 2\otimes \frac 1 2 \otimes
\frac 1 2$ which decomposes as $\frac 3 2 \oplus \frac 1 2 \oplus \frac 1 2$.
Notice that there is no scalar part in this decomposition. This means
that we can not form a term invariant under rotations with three spinors.
This would be needed however to write down an interaction term corresponding
to the Chern-Simons interaction. A similar consideration applies to the
$SU(2)_I$ symmetry. This argument shows that it is impossible to untwist
Chern-Simons theory. Several remarks are in order now.
The possibility of untwisting Chern-Simons based on the twisted $N=4$
algebra has been considered before in \cite{lucchesi}. There the authors
started however with a three dimensional $N=4$ sigma-model with conventional
statistics. The twist had to be supplemented by a change of statistics
of the fields and the substitutions of an $\epsilon_{AB}$ instead of
an $\delta_{AB}$ in the kinetic term of the fermions. As was shown here
it is much more natural to start from the beginning with a sigma-model
of reversed statistics since one automatically deals then with the 
``correct'' kinetic term for the spin $\frac 1 2$ fields. Furthermore,
in \cite{lucchesi} the non-Abelian interaction terms have not been addressed.
On the other hand we could give a fairly general group-theoretical argument 
showing that it is in fact impossible to obtain non-Abelian Chern-Simons 
theory by twisting some kind of sigma-model action. More generally, 
the study of sigma-models with reversed statistics might be an interesting 
area for future research. Of course the actual sigma-model should then not be
taken seriously as a starting point for quantization. What we have in mind
is that the topologically twisted versions of such sigma-models could
give interesting analogs of Chern-Simons theory in three dimensions.

%%%%%%%%%%%%%%%%%%%%%%%%%%%%%%%%%%%%%%%%%%%%%%%%%%
\section{The dimensional reduction to $D=2$}
%%%%%%%%%%%%%%%%%%%%%%%%%%%%%%%%%%%%%%%%%%%%%%%%%%
In this Section we perform a dimensional reduction {\it \`a la} Scherk of 
the algebra presented in Section 2 as well as the Chern-Simons 
model and its symmetries displayed in Section 3. In the latter case, such
a dimensional reduction gives rise to the BF-model in $D=2$ and its
respectives symmetries. A new hidden symmetry of the BF-model 
arises as a direct consequence of the dimensional reduction of the 
topological vector supersymmetry generator.  

\subsection{Dimensional reducing the algebra}
By dimensional reducing the twisted algebra (\ref{tn4}) giving
\be
\{Q_{\,\a}^{\ab~\g},Q_{\,\b}^{\bb~\d}\}=2i\e^{\g\d}\e^{\ab\bb}
\s_{m\a\b}\pa_m ~~,
\label{tn412d}
\ee
($m=1,2$), the operators (\ref{brs}--\ref{topvec}) turn out to be
\letra
\bq
&&s^{\ab}=\left(\begin{array}{c}
s\\
\ats
\end{array}\right)\equiv {1\over{2}}~\d_\a^{~\b}Q_{\,\b}^{\ab~\a}
~~~{\mbox{(2 d.f. from Tr$(Q^{\ab})$)}} ~~;\label{brsbf} \\
&&\d^{\ab}_{m}=\left(\begin{array}{c}
\atd_m\\
\d_m
\end{array}\right)\equiv -{i\over{2}}~\s_{m\a}^{~~\b}Q_{\,\b}^{\ab~\a} 
~~~{\mbox{(4 d.f. from Tr$(\s_m Q^{\ab})$)}}~~;\label{topvecbf} \\
&&\z^{\ab}=\left(\begin{array}{c}
\az\\
\z
\end{array}\right)\equiv -{i\over{2}}~\s_{3\a}^{~~\b}Q_{\,\b}^{\ab~\a} 
~~~{\mbox{(2 d.f. from Tr$(\s_3 Q^{\ab})$)}}~~.\label{extrabf}
\eq
\antiletra 

The operators (\ref{brsbf}--\ref{extrabf}) together with the supersymmetry 
algebra (\ref{tn412d}) yield the following anticommutation relations :
\bq
&&\{s^{\ab},s^{\bb}\}=0 ~~,~~\{s^{\ab},\z^{\bb}\}=0 
\aand 
\{s^{\ab},\d^{\bb}_m\}=-\e^{\ab\bb} \pa_m ~~; \nonumber \\
&&\{\z^{\ab},\z^{\bb}\}=0 ~~,~~\{\z^{\ab},\d^{\bb}_m\}=
\e^{\ab\bb} \ve_{mn}\pa_n
\aand 
\{\d^{\ab}_m,\d^{\bb}_n\}=0 ~~,
\label{alg1bf}
\eq
that lead to the algebra displayed below
\bq
&&\{s,s\}=\{\ats,s\}=\{s,\ats\}=\{\ats,\ats\}=0 ~~;\nonumber \\
&&\{s,\z\}=\{s,\az\}=\{\ats,\z\}=\{\ats,\az\}=0 ~~;\nonumber \\
&&\{s,\d_m\}=\pa_m ~~,~~\{s,\atd_m\}=0 ~~;\nonumber \\
&&\{\ats,\d_m\}=0 ~~,~~\{\ats,\atd_m\}=-\pa_m ~~;\nonumber \\
&&\{\z,\z\}=\{\z,\az\}=\{\az,\z\}=\{\az,\az\}=0 ~~;\nonumber \\
&&\{\z,\d_m\}=0 ~~,~~\{\z,\atd_m\}=\ve_{mn}\pa_n ~~;\nonumber \\
&&\{\az,\d_m\}=-\ve_{mn}\pa_n ~~,~~\{\az,\atd_m\}=0 ~~;\nonumber \\
&&\{\d_m,\d_n\}=\{\d_m,\atd_n\}=\{\atd_m,\d_n\}=\{\atd_m,\atd_n\}=0 ~~.
\label{alg2bf}
\eq

Now, in two dimensions, the supercharges $Q^+$ and $Q^-$ read
\be
Q^+=\I s+i\s_3\az+i\s_m\atd_m \aand Q^-=\I\ats+i\s_3\z+i\s_m\d_m~~,
\label{Q+-2}
\ee
where it can be checked with the help of the algebra (\ref{alg2bf}) 
that $Q^+$ and $Q^-$ satisfy the $N=4$ supersymmetry algebra 
(\ref{tn412d}).

\subsection{The BF-model in $D=2$ and its new symmetries}
The dimensional reduction {\it \`a la} Scherk of the Chern-Simons action 
(\ref{CS}) to $D=2$ is achieved by assuming that
\bq
A_\m(x)~~\dr~~\left(A_m(\hat x),\f(\hat x)\right)~~{\mbox{where}}
&\left\{\begin{array}{l}
x_\m~,~~\m=1,2,3 \\
\\
\hat x_m~,~~m=1,2
~~~~~, \label{dr}
\end{array}\right.
\eq
with $\f$ being a real scalar and $\ve_{mn3}\equiv\ve_{mn}$. Since we are 
adopting the trivial dimensional reduction, the derivative, $\pa_3$, 
of all fields vanishes, $\pa_3{\cf}=0$. Therefore, the following action 
stems 
\be
\S_{{\rm BF}}={k\over 4\p}{\mbox{Tr}}\int d^2{\hat x}~\ve
_{mn}F_{mn}\f ~~,
\label{BF}
\ee
with $F_{mn}=\pa_{[m} A_{n]}+[A_m,A_n]$. 

Now, let us perform the dimensional reduction of the gauge fixing (\ref{gf}), 
by assuming a curved three-dimensional manifold described by the metric 
tensor $g_{\m\n}(x)$ which determinant is denoted by $g$, it follows that
\letra
\bq
&&g_{\m\n}(x)~~\dr~~\left(g_{mn}(\hat x),v_m(\hat x),\vf(\hat x)\right)~~,\\
&&g={\hat g}(\vf-v_mv_m)~~,\label{drg}
\eq
\antiletra 
where ${\hat g}$ is the determinant of the two-dimensional metric $g_{mn}$. 
The reduced gauge fixing in a curved two-dimensional space read
\bq
\S_{{\rm gf}}=
&\left\{\begin{array}{l}
s~{\mbox{Tr}}{\dst\int}d^2{\hat x}~{\sqrt{{\hat g}(\vf-v_kv_k)}}
~g_{mn}~\atc\left(v_m\pa_n\f+\pa_m A_n\right) \\
\\
-\ats~{\mbox{Tr}}{\dst\int}d^2{\hat x}~{\sqrt{{\hat g}(\vf-v_kv_k)}}
~g_{mn}~c\left(v_m\pa_n\f+\pa_m A_n\right) 
~~~~. \label{sgf2}
\end{array}\right.
\eq
Thus, from the energy-momentum tensor (\ref{tmn})  
\bq
T_{\m\n}~~\dr~~\left(T_{mn},T_m,T\right)~~{\mbox{where}}
&\left\{\begin{array}{l}
\left. T_{mn}={2\over{\sqrt g}}
{{\d\S_{\rm gf}}\over{\d g_{mn}}}\right|_{\stackrel{\vf=1}{\sss v_m=0}}=
s\Ld_{mn}=
\ats {\bar\Ld}_{mn} \\
\\
\left. T_{m}={2\over{\sqrt g}}
{{\d\S_{\rm gf}}\over{\d v_m}}\right|_{\stackrel{\vf=1}{\sss v_m=0}}=
s\Ld_{m}=
\ats {\bar\Ld}_{m}\\
\\
\left. T={2\over{\sqrt g}}
{{\d\S_{\rm gf}}\over{\d \vf}}\right|_{\stackrel{\vf=1}{\sss v_m=0}}=0
~~~~, \label{tmn2}
\end{array}\right.
\eq
such that 
\letra
\bq
&&\Ld_{mn}=-{\mbox{Tr}}~\atc\left[\d_{mn}\pa_lA_l - 
(\pa_mA_n+\pa_nA_m)\right]~~, \label{l1}\\
&&{\bar\Ld}_{mn}={\mbox{Tr}}~c\left[\d_{mn}\pa_lA_l - 
(\pa_mA_n+\pa_nA_m)\right]~~, \label{l2}\\
&&\Ld_{m}=2~{\mbox{Tr}}~\atc\pa_m\f~~, \label{l3}\\
&&{\bar\Ld}_{m}=-2~{\mbox{Tr}}~c\pa_m\f~~.\label{l4}
\eq
\antiletra

The action $\S_{{\rm inv}}^{{\rm (2)}}=\S_{{\rm BF}}+\S_{{\rm gf}}$ 
is invariant under the following BRS and \BRS~transformations:
\be
\begin{array}{ll}
sA_m=-D_m c ~~,& \ats A_m=-D_m \atc ~~,\nonumber\\
s\f=-[\f,c] ~~,& \ats\f=-[\f,\atc] ~~,\nonumber\\
sc=c^2 ~~,& \ats\atc=\atc^2 ~~,\nonumber\\
s\atc=b ~~,& \ats c=\{\atc,c\}-b ~~,\nonumber\\
sb=0 ~~,& \ats b=[\atc,b] ~~,\label{transf2} 
\end{array}
\ee
where $D_m\cdot\equiv\pa_m\cdot+[A_m,\cdot]$. From 
(\ref{l1}--\ref{l2}) it
can be found that the action $\S_{{\rm inv}}^{{\rm (2)}}$ is
invariant under the following topological vector supersymmetries :
\be
\begin{array}{ll}
\d_m A_n=0 ~~,& 
\atd_m A_n=0 ~~,\nonumber\\
\d_m \f=-{2\p\over{k}}\ve_{mn}\pa_n \atc ~~,& 
\atd_m \f=-{2\p\over{k}}\ve_{mn}\pa_n c ~~,\nonumber\\
\d_m c=-A_m ~~,& \atd_m\atc=A_m  ~~,\nonumber\\
\d_m b=\pa_m\atc ~~,& \atd_m b=D_m c ~~,\nonumber\\
\d_m\atc=0 ~~,& \atd_m c=0 ~~.\label{vecsusyt2} 
\end{array}
\ee
It should be stressed that beyond the usual topological symmetries 
(\ref{vecsusyt2}) the action $\S_{{\rm inv}}^{{\rm (2)}}$ is also invariant 
under a new hidden ``gauge'' symmetry, a ``topological pseudo-scalar 
supersymmetry'', obtained from (\ref{l3}--\ref{l4})
\be
\begin{array}{ll}
\z A_m={2\p\over{k}}\ve_{mn}\pa_n \atc ~~,& 
\az A_m={2\p\over{k}}\ve_{mn}\pa_n c ~~,\nonumber\\
\z \f=0 ~~,& 
\az \f=0 ~~,\nonumber\\
\z c=-\f ~~,& \az\atc=\f  ~~,\nonumber\\
\z b=0 ~~,& \az b=[\f,c] ~~,\nonumber\\
\z\atc=0 ~~,& \az c=0 ~~.\label{newt2} 
\end{array}
\ee

All the symmetries displayed above in (\ref{transf2}), (\ref{vecsusyt2}) and 
(\ref{newt2}), satisfy the algebra (\ref{alg2bf}) up to contact terms. These 
symmetries of the two-dimensional BF-model $\S_{{\rm inv}}^{{\rm (2)}}$ 
(including the 
new one (\ref{newt2})) could be gotten simply by dimensional reducing the 
symmetries of the original Chern-Simons model $\S_{{\rm inv}}^{{\rm (3)}}$ 
(given by eqs. (\ref{transf}) and (\ref{vecsusyt})).

\subsection{Untwisting the BF-model}
In order to investigate the possibility of untwisting the two-dimensional
BF-model we find it convenient to introduce complex coordinates
\be
\left(
\begin{array}{c} z\\ \bar z \end{array} \right) = \frac 1 2 \left(
\begin{array}{cc} 1&i\\1&-i \end{array} \right) \left(
\begin{array}{c} x^1\\x^2 \end{array} \right)~~.
\ee
The gauge fixed BF-model can then be written as
\be
\S_{{\rm inv}}^{{\rm (2)}}={\mbox{Tr}}\int d^2z~\left(\c\bar{\pa} A_z + 
\bar\c \pa A_{\bar z}
+ {2\p\over{k}}(\bar\c - \c) [A_z,A_{\bar z}] + \atc \pa \bar{\pa} c
+ {\frac 12} \{c,\pa\atc \} A_{\bar z} + {\frac 12} \{c,\bar{\pa}\atc \} 
A_z\right)~~,
\label{bfcomplex}
\ee
where we introduced $\c={\frac b2} + i\frac{k}{4\pi} \f$.
In two dimensions the rotation group is Abelian. The advantage of working
with (\ref{bfcomplex}) is that all the fields have definite spin eigenvalues.
The topological twist consists now of just adding the spin eigenvalues 
and the eigenvalues of the fields under the Cartan generators of the 
$SU(2)_I\otimes SU(2)_{II}$. The question if one can untwist the
BF-model boils down then to the question of which spin eigenvalues one
can assign to the fields. Let us concentrate again for a moment on the
kinetic term alone. Since $(\pa,\bar{\pa})$ have spin $(1,-1)$
and the Lagrangian should of course have spin zero we see that
we can assign spin eigenvalues to $\c$ and $A_z$ such that 
\be
spin(\c) + spin(A_z) = 1~~.
\ee
We also demand $spin(\c)=-spin(\bar{\c})$ and $spin(A_{\bar z})= - spin(A_z)$. 
>From the interaction terms we get the equations
\be
spin(\c) + spin(A_z) + spin(A_{\bar z}) = spin(\bar{\c}) + spin(A_z) + 
spin(A_{\bar z})=0~~.
\ee
These equations together with analogous ones including $(c,\atc)$ tell us 
that the only
possible spin assignments are the ones which we have in the original BF-model
namely $spin(\c,A_z,c,\atc) = (0,1,0,0)$.
If the interaction terms vanish however we can in particular assign
spin $\frac 1 2$ to $(A_z,\c)$. The action is then the one for a 
linear sigma-model with reversed statistics in two dimensions.
Analogous to what we had in the case of the three-dimensional Chern-Simons
theory we find again that we can untwist only the non-interacting
theory.

\section{Conclusions}
We reinvestigated the twisted $N=4$ supersymmetry present in
the three-dimensional Chern-Simons theory and the two-dimensional
BF-model with Landau-type gauge fixing. We could show that 
it is impossible to untwist the non-Abelian theories. For the Abelian
case we also showed that some puzzles concerning the untwisting 
procedure can be avoided by choosing an approach through sigma-models
with reversed statistics in comparison to the G-twist of \cite{lucchesi}.
For the two-dimensional BF-model we established new fermionic (pseudo-)scalar
symmetries ($\z$, $\az$) who together with the BRS, anti-BRS and 
the respective fermionic vector supersymmetries $(\d_\mu, \bar{\d_\m})$ 
give rise to the twisted $N=4$ algebra. We expect these symmetries to be 
useful in the investigation of possible counterterms as {\it e.g.} in 
\cite{schweda}. Finally one could imagine also to apply the methods 
presented here to higher-dimensional analogs of BF- and Chern-Simons models 
as recently investigated in \cite{baulieu}.
 
%%%%%%%%%%%%%%%%%%%%%%%%%%%%%%%%%%
\subsection*{Acknowledgements}
%%%%%%%%%%%%%%%%%%%%%%%%%%%%%%%%%
The authors express their gratitude to Olivier Piguet for his comments and
suggestions.

\newpage

\appendix
\section{Notations and conventions for the Euclidean $D\,$=3}
\setcounter{equation}{0}

Throughout this work the metric is assumed to be Euclidean, 
$\d_{\m\n}=diag(+,+,+)$, $\m$,
$\n$=(1,2,3). The Dirac 2$\times$2 $\g$-matrices are in fact the Pauli 
$\s$-matrices that satisfy the Clifford algebra 
\be
\{\sm,\sn\}=2\d_{\m\n}\I ~~\Longrightarrow~~
\{\sm,\sn\}_\a^{~\b}=2\d_{\m\n}\d_\a^{~\b}~~,
\ee
are as follows
\be
\sm\equiv\s_{\m\a}^{~~\b}=(\s_1,\s_2,\s_3)~~,
\ee
where the Pauli matrices read
\be
\s_1=\left(\begin{array}{cc}0&1\\
                           1&0
\end{array}\right)\;,\;\;
\s_2=\left(\begin{array}{cc}0&-i\\
                           i&0
\end{array}\right)\;,\;\;
\s_3=\left(\begin{array}{cc}1&0\\
                           0&-1
\end{array}\right)\;\;\;.
\ee

Some useful relations involving the $\s$-matrices, their traces (Tr) and
the $\ve$-tensor used in the calculations are given by :
\letra
\bq
&&\sm\sn=\d_{\m\n}\I+i\ve_{\m\n\r}\g_\r~~\Longrightarrow~~
\s_{\m\a}^{~~\b}\s_{\n\b}^{~~\g}=
\d_{\m\n}\d_\a^{~\g}+i\ve_{\m\n\r}\s_{\r\a}^{~~\g}~~; \\
&&\mbox{Tr}(\sm\sn)=2\d_{\m\n}~~\Longrightarrow~~
\s_{\m\a}^{~~\b}\s_{\n\b}^{~~\a}=2\d_{\m\n}~~; \\
&&\ve_{\m\n\k}\ve_{\m\s\r}=\d_{\s\n}\d_{\r\k}-\d_{\r\n}\d_{\s\k}~~.
\eq
\antiletra
The spinor indices are raised and lowered according to the rules 
\be
X^\a=\e^{\a\b}X_\b \aand X_\a=X^\b\e_{\b\a}~~,
\ee
with the $\e_{\a\b}$ tensor defined by
\be
\e_{\a\b}=\left(\begin{array}{cc}0&1\\
                           -1&0
\end{array}\right) \aand
\e^{\a\b}=\left(\begin{array}{cc}0&-1\\
                           1&0
\end{array}\right)~~,
\ee
where the following relation comes
\be
\e^{\a\b}\e_{\g\d}=\d_\g^{~\a}\d_\d^{~\b}-\d_\d^{~\a}\d_\g^{~\b}~~.
\ee

Other important identities follow\footnote{It should be noticed that the 
symmetrizations $(~)$ and antisymmetrizations $[~]$ are just related to 
the spinor indices ($\a,\b,\g,\d$) without a factor $1\over 2$.}
\letra
\bq
&&A_{[\a}^{~[\b}B_{\g]}^{~\d]}=-{1\over 2}\e^{\b\d}
A_{(\a}^{~\e}B_{\g)\e}~~,\\
&&(\I\wedge\sm)_{\a~\g}^{~\b~\d}\equiv 
\d_{[\a}^{~[\b}\s_{\m\g]}^{~~\d]}=-{1\over 2}\e^{\b\d}
\d_{(\a}^{~\e}\s_{\m\g)\e}=-\e^{\b\d}\s_{\m\a\g}~~,
 \\
&&(\sm\wedge\sn)_{\a~\g}^{~\b~\d}\equiv 
 \s_{\m[\a}^{~~[\b}\s_{\n\g]}^{~~\d]}=-{1\over 2}\e^{\b\d}
\s_{\m(\a}^{~~\e}\s_{\n\g)\e}=-i\e^{\b\d}\ve_{\m\n\r}\s_{\r\a\g}~~.
\eq
\antiletra

\end{document}